\documentclass[letterpaper, 10 pt, conference]{ieeeconf}

\usepackage{amsmath}
\IEEEoverridecommandlockouts

\usepackage{url}
\usepackage{amsmath,amssymb,amsfonts}
\usepackage{mathtools}
\usepackage{float}
\usepackage{bm}
\usepackage{booktabs}
\usepackage{multirow}
\title{\LARGE \bf
Residual Bias Compensation Dual Extended Kalman Filter for Physics-Based SOC Estimation in Lithium Iron Phosphate Batteries
}

\author{Feng Guo$^{1,2,3}$, Luis D. Couto$^{2,3}$, Khiem Trad$^{2,3}$, Guangdi Hu$^{4}$ and Mohammadhosein Safari$^{1,3,5}$%
\thanks{$^{*}$This paper has been published in the Proceedings of the 2026 European Control Conference (ECC). Please cite the officially published version available at: \protect\url{https://ieeexplore.ieee.org/document/11625570}. \textcopyright~2026 IEEE. Personal use of this material is permitted. Permission from IEEE must be obtained for all other uses, in any current or future media, including reprinting/republishing this material for advertising or promotional purposes, creating new collective works, for resale or redistribution to servers or lists, or reuse of any copyrighted component of this work in other works.}
\thanks{$^{*}$This work was supported by the Research Foundation - Flanders (FWO) (grant numbers 1252326N).}
\thanks{$^{1}$IUMAT, Hasselt University
(UHasselt), Martelarenlaan 42, 3500, Hasselt, Belgium
        {\tt\small \{feng.guo, momo.safari\}@uhasselt.be}}
\thanks{$^{2}$WET, VITO, Boeretang 200, Mol, 2400, Belgium
        {\tt\small \{luis.coutomendonca, khiem.trad\}@vito.be}}
\thanks{$^{3}$EnergyVille, Thor Park 8310, Genk, 3600, Belgium}
\thanks{$^{4}$School of Vehicles and Intelligent Transportation, Fuyao University of
Science and Technology, Fuzhou, 350109, China
        {\tt\small guangdihu@fyust.edu.cn}}
\thanks{$^{5}$IUMAT, IMEC, 3590, Diepenbeek, Belgium}
}

\begin{document}

\maketitle
\thispagestyle{empty}
\pagestyle{empty}

\begin{abstract}
This paper addresses state-of-charge (SOC) estimation for lithium iron phosphate (LFP) batteries, whose relatively flat open-circuit-voltage (OCV)–SOC characteristic leads to poor observability. A residual bias compensation dual extended Kalman filter (RBC-DEKF) is proposed, in which one EKF estimates the electrochemical states of a control-oriented parameter-grouped single-particle model with thermal effects, while the other estimates a residual bias to correct the voltage observation in real time. By decoupling bias estimation from state estimation, the proposed structure avoids the covariance coupling introduced by bias-augmented single-filter schemes. Validation is conducted on an LFP cell from a public dataset under three representative operating conditions: US06 at 0$^{\circ}$C, DST at 25$^{\circ}$C, and FUDS at 50$^{\circ}$C. Compared with a conventional EKF using the same model and identical state-filter settings, the proposed method reduces the average SOC RMSE from 3.75\% to 0.20\% and the voltage RMSE from 32.8\,mV to 0.8\,mV. The improvement is most evident in the mid-SOC region, demonstrating that residual bias compensation can significantly enhance physics-based SOC estimation for LFP batteries across a wide temperature range.
\end{abstract}

\section{INTRODUCTION}

Lithium ion batteries have become the dominant energy storage technology for electric vehicles and stationary energy systems because of their high energy density, long cycle life, and environmental compatibility \cite{guo2024systematic}. Among various chemistries, lithium iron phosphate (LFP) batteries are particularly favored for their superior safety, thermal stability, and cost-effectiveness \cite{schoberl2024thermal}. However, accurately estimating their internal states such as the state of charge (SOC) is essential to ensure safe operation, extend battery lifespan, and optimize system performance \cite{haghverdi2025review}. The relatively flat open circuit voltage (OCV)-SOC curve of LFP batteries leads to poor observability, which makes reliable state estimation highly challenging, especially in the presence of model uncertainties and measurement noise \cite{che2025enhanced,yan2025breaking}.

Existing approaches for SOC estimation can be broadly classified into three categories \cite{guo2018parameter}. The first category includes direct methods such as the current integration method and the open circuit voltage method. These methods are simple and straightforward, but their estimation errors are usually large and prone to accumulation over time. The second category involves artificial intelligence based methods, which can achieve high accuracy but often suffer from poor generalization, overfitting, and limited interpretability, making them difficult to apply in practical engineering systems. The third category is model based methods, which are the most widely used in practice. Techniques such as the Kalman filter \cite{plett2004extended,bizeray2015lithium} and the particle filter \cite{wang2015method} provide good robustness and interpretability, yet their performance strongly depends on the accuracy of the underlying battery model.

Battery models commonly fall into two categories. The first type is the equivalent circuit model (ECM), which represents the battery output voltage using combinations of electrical components such as voltage sources, resistors, and capacitors. This model is simple in structure, but it does not capture the internal electrochemical mechanisms of the battery. As a result, modeling errors are often difficult to control, particularly under varying temperature conditions, and the model parameters usually need to be obtained through extensive experimental data and lookup tables. The second type is the electrochemical model, also known as the physics-based model, which is derived from the internal electrochemical reactions of lithium ion batteries. This model provides higher accuracy but at the cost of increased complexity. The most widely used example is the pseudo two dimensional (P2D) model \cite{fuller1994simulation}, which has high computational complexity and a complicated structure, making it difficult to use in control system design. In practice, a simplified version called the single particle model (SPM) \cite{guo2025control} is often adopted, which maintains the key physical characteristics while achieving faster computation and a simpler structure.

For LFP batteries, the OCV--SOC curve is relatively flat.
When the battery model contains significant errors, even a small voltage deviation can cause a large SOC estimation error, which may lead to inaccurate or unstable estimation results.
In addition to improving the accuracy of the battery model, another approach is to perform joint estimation of battery parameters and states.
By continuously updating the model parameters in real time, the model accuracy and thus the SOC estimation performance can be improved.
For example, previous work applied dual Kalman filters based on an ECM to achieve high-precision SOC estimation for nickel manganese cobalt (NMC) batteries~\cite{guo2019multi}.
This was possible because the OCV--SOC curve of NMC batteries is steeper, providing better observability.
However, applying the same method to LFP batteries is much more difficult due to their flat OCV--SOC characteristics.
Moreover, existing joint estimation methods still face limitations because the model errors may originate not only from parameter inaccuracies but also from sensor biases, OCV curve mismatches, and unmodeled dynamics.
Adjusting model parameters alone cannot effectively correct all these sources of error.
Yi~\textit{et~al.} treated sensor bias as an additional state variable to compensate for measurement deviations, but their approach was still based on the ECM~\cite{yi2024bias}.

Therefore, this paper proposes a residual bias compensation structure that employs dual extended Kalman filters: one filter estimates the electrochemical states of the battery, while the other independently estimates the residual bias to correct model voltage deviations in real time.
Unlike conventional bias compensation schemes, where the bias is treated as an augmented state within a single filter, the proposed method decouples the bias estimation from the state estimation process.
In traditional bias-augmented EKF formulations, the state vector is expanded to include the bias term, which inevitably increases the filter dimension and introduces strong coupling between the bias dynamics and the battery state covariance.
This coupling can degrade estimation stability and cause numerical divergence, particularly when the model observability is weak, as in LFP batteries with flat OCV--SOC characteristics.
In contrast, the proposed dual-filter structure separates the residual bias update from the electrochemical state update, allowing the bias estimator to directly refine the voltage observation equation without disturbing the internal electrochemical dynamics.
Furthermore, the algorithm is implemented on an electrochemical model, providing improved physical consistency and generalizability compared with ECM-based bias compensation methods.

\section{Electrochemical Model}

In this work, the electrochemical model employed is the SPM, which is formulated based on the control-oriented parameter-grouped single particle model with thermal effects (CPG-SPMT)~\cite{guo2025control,guo2025cpgspmt,guo2025comparative}. Compared with other SPM, the proposed model offers faster computational performance and is expressed in a state-space form that satisfies controllability and observability conditions. The corresponding control-oriented state-space formulation can be written as:
\begin{align}
\label{eq:1}
\dot{\chi}(t) &= \tilde{A}\chi(t) + \tilde{B}u(t), \qquad
\chi(0) = \chi_0, \\
\label{eq:2}
\psi(t) &= \tilde{C}\chi(t) + \tilde{D}u(t),
\end{align}
where the normalized state vector is denoted by
\(\chi = [\chi_p^\top \; \chi_n^\top]^\top\),
and each component \(\chi_i = [\tilde{q}_{1,i} \; \tilde{q}_{2,i}]^\top\) (\(i \in \{p, n\}\)) corresponds to the internal dynamic states of the positive and negative electrodes.
The system excitation is represented by the applied current \(u(t) = I(t)\), while the model output
\(\psi = [\psi_p^\top \; \psi_n^\top]^\top\)
includes the normalized lithium concentrations within each electrode, defined as
\(\psi_i = [\tilde{\overline{c}}_{s,i} \; \tilde{c}_{ss,i}]^\top\),
which denote the average and surface solid-phase concentrations, respectively.

The state matrices of the two electrodes are subsequently arranged in a block-diagonal configuration as:
\begin{equation}
\tilde{A} = \mathrm{diag}(\tilde{A}_p, \tilde{A}_n), \quad
\tilde{B} =
\begin{bmatrix}
\tilde{B}_p^\top & \tilde{B}_n^\top
\end{bmatrix}^\top,
\label{eq:3}
\end{equation}
where
\begin{equation}
\tilde{A}_i =
\begin{bmatrix}
0 & 0\\[3pt]
\dfrac{30}{\alpha_i} & -\dfrac{30}{\alpha_i}
\end{bmatrix}, \quad
\tilde{B}_i =
\begin{bmatrix}
\dfrac{1}{Q_i}\\[3pt]
\dfrac{19}{7Q_i}
\end{bmatrix},
\label{eq:4}
\end{equation}
and the corresponding output matrices are expressed as:
\begin{equation}
\tilde{C} = \mathrm{diag}(\tilde{C}_p, \tilde{C}_n), \quad
\tilde{D} =
\begin{bmatrix}
\tilde{D}_p^\top & \tilde{D}_n^\top
\end{bmatrix}^\top,
\label{eq:5}
\end{equation}
with
\begin{equation}
\tilde{C}_i =
\begin{bmatrix}
1 & 0\\
0 & 1
\end{bmatrix}, \quad
\tilde{D}_i =
\begin{bmatrix}
0\\[3pt]
\dfrac{\alpha_i}{105Q_i}
\end{bmatrix}.
\label{eq:6}
\end{equation}

Here, $\alpha_i$ denotes the solid-phase diffusion time constant and $Q_i$ the electrode capacity.
The terminal voltage predicted by the SPM is formulated as:
\begin{equation}
\begin{aligned}
V_{\mathrm{SPM}}(t) &=
{\rm OCP}_p\bigl(\tilde{c}_{ss,p}(t)\bigr)
- {\rm OCP}_n\bigl(\tilde{c}_{ss,n}(t)\bigr) \\
&\quad + \eta_p\bigl(\tilde{c}_{ss,p}(t), I(t)\bigr)
- \eta_n\bigl(\tilde{c}_{ss,n}(t), I(t)\bigr) \\
&- R_0 I(t),
\end{aligned}
\label{eq:8}
\end{equation}
where ${\rm OCP}_i(\tilde{c}_{ss,i})$ is the open-circuit potential of electrode~$i$, and $R_0$ represents the ohmic resistance. The overall open-circuit voltage (OCV) of the cell equals the difference between the positive and negative electrode OCPs, that is, $\mathrm{OCV} = {\rm OCP}_p - {\rm OCP}_n$.
The electrode overpotential is described by:
\begin{equation}
\begin{split}
\eta_{i}(\tilde{c}_{ss,i}(t), I(t))
&= {} \\[-2pt]
&\hspace{-2em}\dfrac{2RT}{F}
\sinh^{-1}\!\left(
\dfrac{1_{\mp}I(t)}
{6Q_i d_i\sqrt{\tilde{c}_{ss,i}(t)\bigl(1-\tilde{c}_{ss,i}(t)\bigr)}}
\right),
\end{split}
\label{eq:7}
\end{equation}
where $d_i$
is the inverse of the reaction time scale of electrode~$i$
and $1_{\mp}$ represents the current direction sign for each electrode ($-1$ and $+1$ for the positive and negative electrodes, respectively).

Each temperature-dependent parameter is expressed relative to its reference value (subscript~1) at $T_{\mathrm{ref}} = 298.15\,\mathrm{K}$ (25$^{\circ}$C):
\begin{equation}
\alpha_n(T) =
\dfrac{\alpha_{n,1}}{\exp\!\left(
\dfrac{E_1}{R}
\left(\dfrac{1}{T_{\mathrm{ref}}} - \dfrac{1}{T}\right)
\right)},
\label{eq:9}
\end{equation}

\begin{equation}
\alpha_p(T) =
\dfrac{\alpha_{p,1}}{\exp\!\left(
\dfrac{E_2}{R}
\left(\dfrac{1}{T_{\mathrm{ref}}} - \dfrac{1}{T}\right)
\right)},
\label{eq:10}
\end{equation}

\begin{equation}
d_n(T) =
d_{n,1}\exp\!\left(
\dfrac{E_3}{R}
\left(\dfrac{1}{T_{\mathrm{ref}}} - \dfrac{1}{T}\right)
\right),
\label{eq:11}
\end{equation}

\begin{equation}
d_p(T) =
d_{p,1}\exp\!\left(
\dfrac{E_4}{R}
\left(\dfrac{1}{T_{\mathrm{ref}}} - \dfrac{1}{T}\right)
\right),
\label{eq:12}
\end{equation}

\begin{equation}
R_{0}(T) =
\dfrac{R_{0,1}}{\exp\!\left(
\dfrac{E_5}{R}
\left(\dfrac{1}{T_{\mathrm{ref}}} - \dfrac{1}{T}\right)
\right)}.
\label{eq:13}
\end{equation}

Here, $T$ denotes the cell temperature, $R$ is the universal gas constant, and $E_1$ to $E_5$ represent activation energies
associated with diffusion, charge-transfer kinetics, and ohmic resistance.

\section{Residual Bias Compensation Dual Extended Kalman Filter (RBC-DEKF)}

Based on the zero-order hold (ZOH) discretization of the CPG-SPMT model in
Eqs.~\eqref{eq:1} and~\eqref{eq:2}, the discrete-time state-space representation is written as
\begin{align}
    \chi_{k+1} &= A_d\,\chi_k + B_d\,u_k + w_k,
    \qquad w_k \sim \mathcal{N}(0,Q_x), \label{eq:14}\\
    \psi_k     &= C_d\,\chi_k + D_d\,u_k, \label{eq:15}
\end{align}
where the electrochemical state vector and the output vector are defined as
\begin{align}
\chi_k &= [\tilde q_{1,p,k},\tilde q_{2,p,k},\tilde q_{1,n,k},\tilde q_{2,n,k}]^\top, \label{eq:16}\\
\psi_k &= [\tilde{\overline c}_{s,p,k},\tilde c_{ss,p,k},\tilde{\overline c}_{s,n,k},\tilde c_{ss,n,k}]^\top, \label{eq:17}
\end{align}
and $w_k$ is a process-noise sequence with covariance matrix $Q_x$. Using $\psi_k$, in particular the surface concentrations $\tilde c_{ss,i}$, the SPM-predicted voltage is obtained from Eq.~\eqref{eq:8} and compactly denoted as
\begin{equation}
    V_{\mathrm{SPM}}(k) = \mathcal{V}\big(\psi_k,u_k\big).
\label{eq:18}
\end{equation}

To account for model mismatch, sensor bias, and unmodeled dynamics, a residual bias state $\theta_k$ is introduced into the voltage measurement equation:
\begin{align}
    \theta_{k+1} &= \theta_k + w_{\theta,k},
    \qquad w_{\theta,k} \sim \mathcal{N}(0,Q_\theta), \label{eq:19}\\
    z_k &= h(\chi_k,u_k,\theta_k) + v_k \notag\\
        &= \mathcal{V}(\psi_k,u_k) + \theta_k + v_k,
    \qquad v_k \sim \mathcal{N}(0,R_x), \label{eq:20}
\end{align}
where $z_k$ denotes the measured terminal voltage, $v_k$ is a measurement-noise sequence with covariance matrix $R_x$, and $w_{\theta,k}$ is a white-noise sequence associated with the residual bias variation, with covariance matrix $Q_\theta$. Both the state filter and the residual-bias filter use the same voltage measurement equation in Eq.~\eqref{eq:20}. However, different measurement-noise covariance settings, namely $R_x$ and $R_\theta$, are adopted for the two filters to allow independent covariance tuning.

At each sampling instant, two coupled filters are executed sequentially. The first EKF estimates the electrochemical states $\chi$, and the second scalar EKF updates the residual bias $\theta$. The coupling arises from the common voltage observation and the latest counterpart estimate.

For the EKF focusing on electrochemical state estimation, the prediction step is
\begin{align}
    \hat{\chi}_{k|k-1} &= A_d \hat{\chi}_{k-1|k-1} + B_d u_{k-1}, \label{eq:21}\\
    P^x_{k|k-1} &= A_d P^x_{k-1|k-1} A_d^\top + Q_x, \label{eq:22}
\end{align}
where $P^x$ is the state estimation error covariance matrix. The corresponding output and voltage predictions are
\begin{align}
    \hat{\psi}_{k|k-1} &= C_d \hat{\chi}_{k|k-1} + D_d u_{k-1}, \label{eq:23}\\
    \hat V_{{\rm SPM},k} &= \mathcal{V}\big(\hat{\psi}_{k|k-1}, u_{k-1}\big). \label{eq:24}
\end{align}

The innovation, Kalman gain, and update step, using the latest bias estimate $\hat{\theta}_{k-1|k-1}$, are given by
\begin{align}
    y^x_k &= z_k - \big(\hat V_{{\rm SPM},k} + \hat{\theta}_{k-1|k-1}\big), \label{eq:25}\\
    S^x_k &= H^x_k\,P^x_{k|k-1}\,(H^x_k)^\top + R_x, \label{eq:26}\\
    K^x_k &= P^x_{k|k-1}\,(H^x_k)^\top\,\big(S^x_k\big)^{-1}, \label{eq:27}\\
    \hat{\chi}_{k|k} &= \hat{\chi}_{k|k-1} + K^x_k\,y^x_k, \label{eq:28}\\
    P^x_{k|k} &= (I - K^x_k H^x_k)\,P^x_{k|k-1}\,(I - K^x_k H^x_k)^\top
    + K^x_k R_x (K^x_k)^\top, \label{eq:29}
\end{align}
where $I$ is the identity matrix, $H^x_k$ is the Jacobian of the measurement equation with respect to the state, and $K^x_k$ is the Kalman gain. The covariance update adopts the Joseph stabilized form to ensure numerical consistency.

For the EKF focusing on residual bias compensation, the prediction step is
\begin{align}
    \hat{\theta}_{k|k-1} &= \hat{\theta}_{k-1|k-1}, \label{eq:30}\\
    P^\theta_{k|k-1} &= P^\theta_{k-1|k-1} + Q_\theta, \label{eq:31}
\end{align}
where $P^\theta$ is the residual-bias estimation error covariance matrix. Using the latest state estimate $\hat{\chi}_{k|k}$, the corresponding output and voltage predictions are
\begin{align}
    \hat{\psi}_{k|k} &= C_d \hat{\chi}_{k|k} + D_d u_{k-1}, \label{eq:32}\\
    \hat V_{{\rm SPM},k}^{\,x} &= \mathcal{V}\big(\hat{\psi}_{k|k}, u_{k-1}\big). \label{eq:33}
\end{align}

The innovation, gain, and update are then given by
\begin{align}
    y^\theta_k &= z_k - \big(\hat V_{{\rm SPM},k}^{\,x} + \hat{\theta}_{k|k-1}\big), \label{eq:34}\\
    S^\theta_k &= H^\theta_k\,P^\theta_{k|k-1}\,(H^\theta_k)^\top + R_\theta, \label{eq:35}\\
    K^\theta_k &= P^\theta_{k|k-1}\,(H^\theta_k)^\top\,\big(S^\theta_k\big)^{-1}, \label{eq:36}\\
    \hat{\theta}_{k|k} &= \hat{\theta}_{k|k-1} + K^\theta_k\,y^\theta_k, \label{eq:37}\\
    P^\theta_{k|k} &= (I - K^\theta_k H^\theta_k)\,
    P^\theta_{k|k-1}\,(I - K^\theta_k H^\theta_k)^\top
    + K^\theta_k R_\theta (K^\theta_k)^\top. \label{eq:38}
\end{align}
where $H^\theta_k$ is the Jacobian of the measurement equation with respect to the residual bias, and $K^\theta_k$ is the Kalman gain.

The step-by-step implementation of the RBC-DEKF is summarized as follows:
\begin{enumerate}
    \item State-filter prediction and voltage prediction: Eqs.~\eqref{eq:21}--\eqref{eq:24}.
    \item State-filter update using $\hat{\theta}_{k-1|k-1}$: Eqs.~\eqref{eq:25}--\eqref{eq:29}.
    \item Residual-bias filter prediction and voltage prediction using $\hat{\chi}_{k|k}$: Eqs.~\eqref{eq:30}--\eqref{eq:33}.
    \item Residual-bias filter update: Eqs.~\eqref{eq:34}--\eqref{eq:38}.
\end{enumerate}

The shared measurement function is written as
\begin{equation}
h(\chi,u,\theta)=\mathcal{V}(\psi,u)+\theta,
\qquad
\psi=C_d\chi+D_du.
\label{eq:39}
\end{equation}
Equation~\eqref{eq:39} indicates that the predicted terminal voltage is composed of two parts: the model-based voltage $\mathcal{V}(\psi,u)$ and the additive residual bias term $\theta$.

By the chain rule, the Jacobian with respect to the state is
\begin{equation}
H^x_k=\frac{\partial h}{\partial \chi}\Big|_{\hat\chi,\hat\theta,u}
=\frac{\partial \mathcal{V}}{\partial \psi}\Big|_{\hat\psi,u}\, C_d,
\label{eq:40}
\end{equation}
while the Jacobian with respect to the residual bias is
\begin{equation}
H^\theta_k
= \frac{\partial h}{\partial \theta}\Big|_{\hat\chi,\hat\theta,u}
= \frac{\partial \mathcal{V}}{\partial \psi}
   \frac{\partial \psi}{\partial \theta}
  + \frac{\partial \mathcal{V}}{\partial \theta}
  + \frac{\partial \theta}{\partial \theta}.
\label{eq:41}
\end{equation}
 Here, the first two terms vanish because the residual bias~$\theta$ does not affect the internal state vector~$\chi$ or the voltage model
$\mathcal{V}(\psi,u)$, i.e.,
$\partial \psi / \partial \theta = 0$ and $\partial \mathcal{V} / \partial \theta = 0$.
Only the direct additive term in the measurement equation contributes
($\partial \theta / \partial \theta = 1$),
yielding $H^\theta = 1$.
This indicates that the residual bias does not participate in the state dynamics or alter the modeled voltage itself; instead, it acts only as an additive correction term in the measurement function.

After each iteration of the state filter, the estimated state vector is used to reconstruct the normalized lithium concentrations within each electrode. The electrode-level SOC is then obtained by normalizing the estimated average solid-phase concentration with respect to its stoichiometric limits:
\begin{equation}
\mathrm{SOC}_{i,k} =
\frac{\tilde{\overline{c}}_{s,i,k} - \tilde{\overline{c}}_{s,i,\min}}
{\tilde{\overline{c}}_{s,i,\max} - \tilde{\overline{c}}_{s,i,\min}},
\qquad i \in \{p,n\},
\label{eq:42}
\end{equation}
where $\tilde{\overline{c}}_{s,i,\min}$ and $\tilde{\overline{c}}_{s,i,\max}$ denote the lower and upper stoichiometric limits of electrode $i$, respectively. Finally, the overall cell SOC is calculated as the mean of the positive- and negative-electrode SOCs:
\begin{equation}
\mathrm{SOC}_k =
\frac{1}{2}\Big(\mathrm{SOC}_{p,k} + \mathrm{SOC}_{n,k}\Big).
\label{eq:43}
\end{equation}

\section{Results and Discussion}

\begin{figure}[!h]
\centering
\includegraphics[width=1\linewidth]{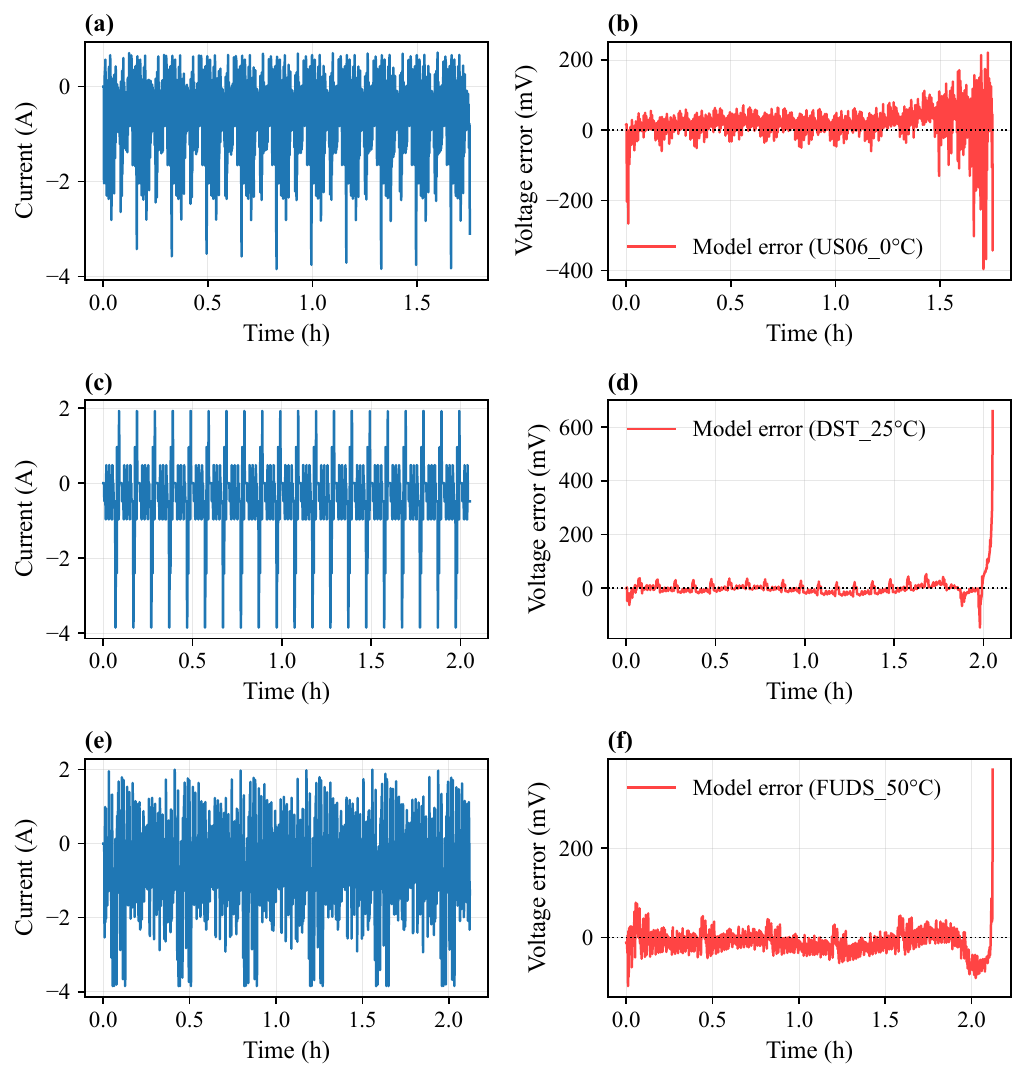}
\caption{Experimental validation of the CPG-SPMT battery model under different operating conditions. The figure shows current profiles (left panels) and voltage error between measurements and CPG-SPMT model predictions (right panels) for three test conditions: (a,b) US06  at 0°C, (c,d) DST at 25°C, and (e,f) FUDS at 50°C. }
\label{fig:1}
\end{figure}

The proposed RBC-DEKF was evaluated using an A123~LFP~18650 cell from the publicly available CALCE battery dataset~\cite{calce2025dataset}.
The CPG-SPMT parameters were identified offline using particle swarm optimization (PSO), and the identification procedure as well as the complete model parameterization follow our previous work in~\cite{guo2025cpgspmt}.
The complete test set covers DST, FUDS, and US06 dynamic current profiles over a temperature range from $-10\,^{\circ}\mathrm{C}$ to $50\,^{\circ}\mathrm{C}$, yielding 24 operating conditions in total.
To present the results concisely in this conference paper, three representative operating conditions were selected for validation and discussion: US06 at 0$^{\circ}$C, DST at 25$^{\circ}$C, and FUDS at 50$^{\circ}$C.
The voltage RMSE of the CPG-SPMT model without filtering was 42.0~mV for US06 at 0$^{\circ}$C, 32.1~mV for DST at 25$^{\circ}$C, and 26.2~mV for FUDS at 50$^{\circ}$C, respectively.
The experimental validation results of the CPG-SPMT model under different temperatures and current profiles are shown in Fig.~\ref{fig:1}.

These open-loop model errors provide a reference for evaluating the subsequent filtering performance of the EKF and the proposed RBC-DEKF. A conventional EKF using the same CPG-SPMT model served as the baseline for comparison.
The process noise covariance, measurement noise covariance, and initial error covariance of the baseline EKF and the electrochemical state filter in the proposed RBC-DEKF were kept identical, ensuring that the observed performance differences arise solely from the inclusion of the residual bias compensation filter. These filter tuning quantities were selected empirically to achieve stable convergence across all operating conditions. A separate covariance configuration was adopted for the residual-bias filter to capture slowly varying voltage residuals. All filter settings were fixed across all test conditions.

\begin{figure}[!h]
\centering
\includegraphics[width=1\linewidth]{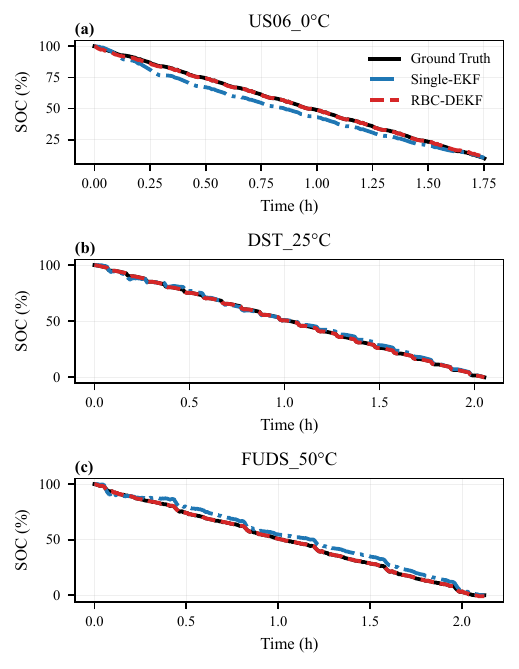}
\caption{SOC estimation comparison between single EKF and RBC-DEKF under: (a) US06 at 0$^{\circ}$C, (b) DST at 25$^{\circ}$C, and (c) FUDS at 50$^{\circ}$C.}
\label{fig:2}
\end{figure}

\begin{figure}[!h]
\centering
\includegraphics[width=1\linewidth]{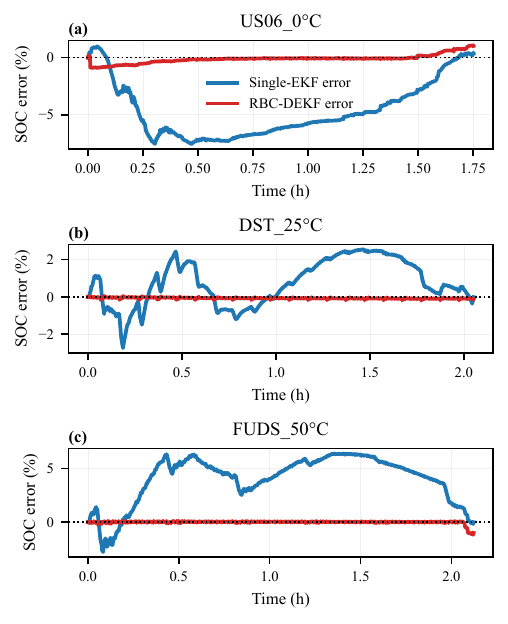}
\caption{SOC estimation error comparison under: (a) US06 at 0$^{\circ}$C, (b) DST at 25$^{\circ}$C, and (c) FUDS at 50$^{\circ}$C.}
\label{fig:3}
\end{figure}

Figs.~\ref{fig:2} and~\ref{fig:3} jointly illustrate the SOC estimation trajectories and corresponding errors of the proposed RBC-DEKF and the baseline EKF under three representative operating conditions.

A consistent pattern is observed across all tests: the RBC-DEKF yields SOC estimates that closely follow the reference SOC with negligible bias, while the EKF accumulates observable drift or transient deviations depending on temperature and load dynamics. The reference SOC, calculated by Coulomb counting from a known initial condition, serves as the benchmark for evaluation.

Under the US06 profile at 0$^{\circ}$C, the battery model error increases due to low-temperature effects.
The baseline EKF exhibits a noticeable drift in the mid-SOC region, while it converges effectively toward the true value at both the fully charged and discharged ends (Fig.~\ref{fig:3}(a)).
Quantitatively, the EKF yields an SOC RMSE of 5.23\%, with a maximum deviation exceeding 7\%.
This behavior occurs because the OCV–SOC curve of LFP batteries is extremely flat in the mid-SOC range, which reduces the voltage sensitivity to SOC and makes accurate estimation difficult when model mismatch is significant.
In contrast, the proposed RBC-DEKF maintains high accuracy across the entire SOC range, with an RMSE of only 0.38\%, particularly improving performance in the mid-SOC region.
This demonstrates that the residual bias filter effectively compensates for model discrepancies and substantially enhances SOC estimation accuracy under low-temperature conditions.

At 25$^{\circ}$C (DST profile), both algorithms perform reasonably well because the battery model accurately captures the electrochemical dynamics at room temperature.
The EKF yields an SOC RMSE of 1.43\%, with instantaneous errors oscillating between approximately $-2\%$ and $+2\%$ following each current pulse (Fig.~\ref{fig:3}(b)), reflecting transient mismatch during step changes.
In contrast, the RBC-DEKF effectively suppresses these oscillations and maintains an SOC RMSE of only 0.08\%, producing trajectories that nearly coincide with the reference in Fig.~\ref{fig:2}(b).

Under the FUDS profile at 50$^{\circ}$C, the baseline EKF exhibits noticeable drift in the mid-SOC region, with an RMSE of 4.59\% and a maximum deviation exceeding 5\% (see Fig.~\ref{fig:3}(c)).
In contrast, the proposed RBC-DEKF maintains the SOC error tightly bounded around zero, achieving an RMSE of only 0.14\%.
These results confirm that the proposed method remains effective under high-temperature conditions, demonstrating strong robustness and adaptability in handling temperature-dependent nonlinearities and model uncertainties. In summary, Figs.~\ref{fig:2} and~\ref{fig:3} demonstrate that the proposed RBC-DEKF consistently achieves accurate SOC estimation across a wide temperature range.

\begin{figure}[!h]
\centering
\includegraphics[width=1\linewidth]{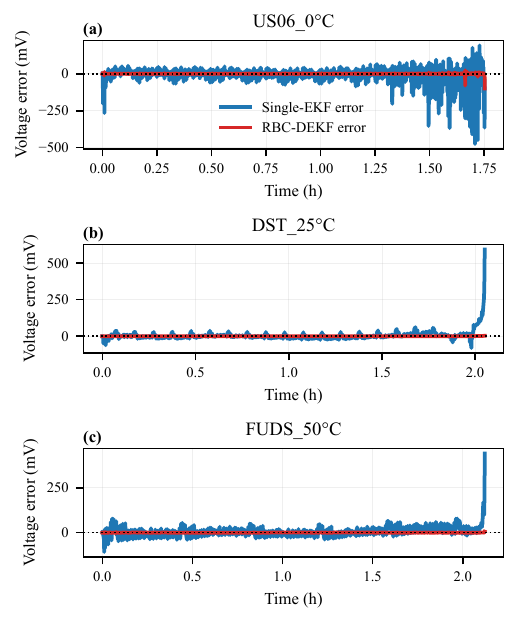}
\caption{Voltage estimation error comparison under: (a) US06 at 0$^{\circ}$C, (b) DST at 25$^{\circ}$C, and (c) FUDS at 50$^{\circ}$C.}
\label{fig:4}
\end{figure}

Fig.~\ref{fig:4} shows the voltage error comparison.
The baseline EKF exhibits a clear offset between model-predicted and reference voltage, whereas the RBC-DEKF successfully removes this offset and significantly narrows the error band. This confirms that continuously estimating the residual bias helps reconcile the difference between modeled and actual voltage behavior of LFP cells, thereby improving the accuracy of the subsequent SOC estimation.

Table~\ref{tab:performance} summarizes the quantitative results. It should be noted that the voltage RMSE values reported in Table~\ref{tab:performance} correspond to the voltage residual after filtering, rather than the open-loop model error shown in Fig.~\ref{fig:1}.
Across all conditions, the proposed RBC-DEKF achieves an average SOC~RMSE of 0.20\%, corresponding to a 94.7\% improvement over the single EKF.
Voltage estimation accuracy also improves substantially, with RMSE reduced from 32.8~mV to 0.8~mV (a 97.5\% reduction).
It should be noted that the residual-bias compensation does not change the intrinsic flatness of the OCV--SOC curve. Instead, it reduces the systematic voltage mismatch seen by the state filter, thereby preventing the EKF from interpreting model-induced voltage residuals as SOC correction signals, which is particularly important in the mid-SOC region of LFP cells.

\begin{table}[!h]
\centering
\caption{Performance Comparison Between EKF and RBC-DEKF}
\label{tab:performance}
\renewcommand{\arraystretch}{1.2}
\setlength{\tabcolsep}{3pt}
\begin{tabular}{lccccccc}
\toprule
\multirow{2}{*}{Condition}
 & \multicolumn{2}{c}{SOC RMSE (\%)}
 & \multicolumn{2}{c}{V RMSE (mV)}
 & \multicolumn{2}{c}{Improvement (\%)}\\
\cmidrule(lr){2-3}\cmidrule(lr){4-5}\cmidrule(lr){6-7}
 & EKF & RBC-DEKF & EKF & RBC-DEKF & SOC & Voltage\\
\midrule
US06 (0$^{\circ}$C) & 5.23 & \textbf{0.38} & 45.3 & \textbf{2.0} & 92.7 & 95.6\\
DST (25$^{\circ}$C) & 1.43 & \textbf{0.08} & 30.0 & \textbf{0.1} & 94.7 & 99.7\\
FUDS (50$^{\circ}$C) & 4.59 & \textbf{0.14} & 23.1 & \textbf{0.3} & 96.9 & 98.7\\
\midrule
\textbf{Average} & 3.75 & \textbf{0.20} & 32.8 & \textbf{0.8} & \textbf{94.7} & \textbf{97.5}\\
\bottomrule
\end{tabular}
\end{table}

Overall, the results show that estimating and compensating the model residual through a dual-EKF architecture substantially enhances SOC accuracy for LFP cells.

\section{CONCLUSIONS}

This paper proposed an RBC-DEKF for accurate SOC estimation of LFP batteries.
The method is built upon an electrochemical model and employs two coupled EKFs that share a common voltage observation. The state EKF estimates the system states of the electrochemical model, while the residual bias EKF estimates a residual term that continuously corrects the voltage observation equation to account for model discrepancies in real time.
Experimental validation using an LFP cell demonstrated substantial improvements over a conventional EKF configured with identical state-filter parameters.
On average, the proposed RBC-DEKF reduced the SOC RMSE from 3.75\% to 0.20\% and the voltage RMSE between the voltage residual after filtering and the measured voltage from 32.8\,mV to 0.8\,mV.
The filter maintained precise tracking across wide temperature ranges and effectively mitigated the mid-SOC drift caused by the flat OCV–SOC curve of LFP chemistry.
Future work will extend the proposed framework to jointly estimate other parameters, as well as the theoretical proof and experimental verification of filter stability and robustness.

\addtolength{\textheight}{-12cm}

\end{document}